\documentclass[a4paper,12pt]{article}

\usepackage[T1]{fontenc} 
\usepackage[swedish,american]{babel}

\usepackage[backend=bibtex,style=numeric]{biblatex} 
\DeclareLanguageMapping{american}{american-apa}
\let\cite\parencite
\let\citeN\textcite

\usepackage{hyperref}
\usepackage{comment}
\usepackage{url}
\usepackage{amssymb}                             
\usepackage{paralist}
\usepackage{xspace}
\usepackage[shadow,textsize=small]{todonotes}

\usepackage{vmargin} 
\setpapersize{A4}
\setmarginsrb{1in}{20pt}{1in}{.5in}{1pt}{2pt}{0pt}{2mm}

\newcommand{\ie}{\textit{i.e.,}\xspace}
\newcommand{\eg}{\textit{e.g.,}\xspace}

\usepackage[normalem]{ulem}

\begin{document}


\title{Comparing Properties of Massively Multiplayer Online Worlds and the Internet of Things [v-0.9]}
\author{Kim J.L. Nevelsteen, Theo Kanter, Rahim Rahmani}
\date{}
\maketitle

\newcommand{\DOMAIN}{}
\newcommand{\RQ}{}
\newcommand{\PROBLEM}{}
\newcommand{\RELATEDWORK}{}
\newcommand{\APPROACH}{}
\newcommand{\VERIFICATION}{}
\newcommand{\RESULTS}{}
\newcommand{\IMPLICATIONS}{}

\newcommand{\IoT}{IoT\xspace}
\newcommand{\WSAN}{WSAN\xspace}
\newcommand{\MMO}{MMO\xspace}
\newcommand{\MMOW}{MMOW\xspace}
\newcommand{\PtoP}{P2P\xspace}

\newcommand{\KEYW}[1]{\mbox{\textbf{#1}}\xspace}

\begin{abstract}

\PROBLEM
With the rise of the Internet of Things (\IoT), this means recognizing the need for architectures to handle billions of devices and their interactions. 
\DOMAIN 
A virtual world engine at the massively multiplayer scale is a massively multiplayer online world (\MMOW); one thing virtual world engines realized when going into the scale of {\MMO}s, is the cost of maintaining a potentially quadratic number of interactions between a massive number of objects, laid out in a spatial dimension. 
\APPROACH
Research into \IoT was fueled by research in wireless sensor networks, but rather than start from a device perspective, this article looks at how architectures deal with interacting entities at large scale. 
\VERIFICATION
The domain of {\MMOW}s is examined for properties that are affected by scale. Thereafter the domain of \IoT is evaluated to see if each of those properties are found and how each is handled.  
\RESULTS
By comparing the current state of the art of {\MMOW}s and \IoT, with respect to scalability, the problem of scaling \IoT is explicated, as well as the problem of incorporating an \MMOW with \IoT into a pervasive platform.
Three case studies of a \MMOW interfacing with \IoT are presented in closing.

\end{abstract}


\section{Introduction}

\DOMAIN 
A virtual world at the massively multiplayer scale is a Massively Multiplayer Online World (\MMOW)~\cite{nevelsteenDRAFT-virtualworlddef}. 
One thing virtual world engines realized when going into the scale of {\MMO}s, is the cost of maintaining a potentially quadratic number of interactions between a massive number of objects, laid out in a spatial dimension~\mbox{\cite{yahyavi2013-p2pmmog,hu2006-von}}. \citeN{yahyavi2013-p2pmmog} explicitly state that the architectures they focus on for \MMO games are also applicable to other distributed systems \eg technology-sustained pervasive games~\cite{nevelsteen2015-pervasivemoo}.

\PROBLEM
\RELATEDWORK
With the rise of the Internet of Things (\IoT)~\cite{gubbi2013-vision,perera2014-contextaware,atzori2010-iot}, this means recognizing the need for architectures to handle billions of devices and their interactions. 
Endeavors are already underway in an attempt to create an \IoT platform, but a ``solution that addresses all the aspects required by the \IoT is yet to be designed''~\cite{perera2014-contextaware}. \citeN{miorandi2012-adhoc} summarize a number of research initiatives happening worldwide \eg HYDRA allowing developers to incorporate heterogeneous devices and IoT-A concentrating on interoperability.
Devices in \IoT (\eg RFID) allow for the mapping of the physical world into the virtual world~\cite{atzori2010-iot} \ie using `non-standard input devices' to blend the virtual and the physical~\cite{nevelsteen2015-pervasivemoo} into a pervasive system~\cite{rehm2015-mediator}.

\APPROACH
\VERIFICATION
Research into \IoT was fueled by research in Wireless Sensor and Actuator Networks ({\WSAN}s)~\cite{gubbi2013-vision}, but rather than start from a device perspective, in Section~\ref{section:approach}, the domain of {\MMOW}s is examined to see how architectures deal with scalability \ie properties that are affected by scale are gathered from the domain of {\MMOW}s.
In Section~\ref{section:evaluation}, it is then evaluated if each of those properties is found in the domain of \IoT and how each property is handled. By comparing the current state of the art of {\MMOW}s and \IoT, with respect to scalability, Section~\ref{section:discussion} points to how research from one domain can possibly be exapted to the other domain and \textit{vice versa}. In addendum, Section~\ref{section:casestudies} provides three case studies as to how an \MMOW can interface with \IoT or how a \MMOW can be a `mediator'~\cite{rehm2015-mediator} for \IoT. The article closes with conclusions summarized in Section~\ref{section:conclusions}.

\section{Examining the Domain of {\MMOW}s}
\label{section:approach}

Since virtual world engines (implementing {\MMOW}s) have long been dealing with the cost of maintaining a massive number of objects, issues are gathered from the domain of {\MMOW}s that are affected by scale. Issues are categorized according to properties presented in the ISO 25010:2011 standard~\cite{iso2011-25010}. %

An issue, relative to {\MMOW}s, that is currently under research is that of \KEYW{scalability}. ``Scalability can be achieved either by:%
\begin{inparaenum}[(1)]
\item increasing the resources or by 
\item reducing the consumption%
\end{inparaenum} 
''~\cite{yahyavi2013-p2pmmog}.
Adding multiple servers to distribute the load of handling interactions is designed to increase the amount of resources. 
To alleviate the scalability problem, a server cluster can be used instead of the single server \ie a co-located cluster of servers that collectively act as a centralized unit to serve all clients. To resolve the scalability issue further, current MMO research is looking into pure peer-to-peer (\PtoP) solutions or a hybrid \PtoP server cluster combination. Server clusters can be formed in a distributed \PtoP system by assigning a `region controller'~\cite{hampel2006-p2pmmog} to supervise over the peers in the cluster.
If multiple servers are used, two ways to partition computational space are regions and shards \ie regionalization and replication. Regionalization divides space into regions, with a different set of servers responsible for a different region, and replication means having multiple copies of the same space. With replicated shards there is no or minimal interaction between shards~\cite{yahyavi2013-p2pmmog}. A hybrid is also possible \eg a shard divided into regions.
To achieve scalability through a decrease of consumption, `interest management' can be used in combination with partitioning; the amount of resources an entity consumes is limited by assigning each entity an `area of interest'. There are many ways to perform interest management, both structured and unstructured, and several challenges remain (the reader is guided to\citeN{yahyavi2013-p2pmmog} for details).

Latency can be defined as ``the delay between execution of an update at the primary copy of an object and the replica receiving the object update"~\cite{yahyavi2013-p2pmmog}. 
One of the critical aspects of {\MMOW}s is their real-time \KEYW{responsiveness}, which demands ``message latency should be minimized while bandwidth use should be efficient''~\cite{hu2006-von}. From a human perspective, real-time responsiveness is achieved when ``the time between the event being generated and the time it is executed and perceived by the user is unperceivable [sic]''~\cite{liu2012-melding} \ie responsiveness despite latency. %
The tolerance threshold for latency in games is between 100 and 300 milliseconds, depending on the game type (ranging from FPS games to RPGs)~\cite{yahyavi2013-p2pmmog}. 
\citeN{liu2012-melding} report a trade-off between consistency and responsiveness (throughput). %

If multiple servers are collaborating to maintain a compute space, then \KEYW{consistency} between the state of each server must be maintained despite networking delay. \citeN{yahyavi2013-p2pmmog} report a well-known trade-off between performance (availability) and consistency restrictions. Consistency involves having a primary copy of each object and sending updates (\ie update dissemination) to replicas, %
in such a way that the causal order of events are consistent~\cite{yahyavi2013-p2pmmog,liu2012-melding}. Note, the difference between objects being replicated here, and entire server partitions being replicated as with shards. A way to assess a degree of inconsistency is by comparing the (potentially inconsistent) state of each replica against a virtual perfect replica~\cite{yahyavi2013-p2pmmog}.

Considering the prevalence of smartphones, some {\MMOW}s have moved a mobile platform. But the limitations in networking and computing power of such devices are a factor~\cite{yahyavi2013-p2pmmog}. Taking into account whether to use local or remote resources, is important for efficient \KEYW{resource utilization}.

For {\MMOW}s using a centralized cluster of servers, the \KEYW{availability} of end nodes in the network is not so much an issue, but in \PtoP, nodes are ``much more prone to failures or unscheduled disconnections''~\cite{yahyavi2013-p2pmmog}, which can adversely affect the network topology~\cite{hu2006-von}. Related to the availability of the network, nodes must be fault tolerant and support data persistence for recoverability. Data persistence means the ability to save and access world states despite disconnections, which remains a challenge in \PtoP systems~\cite{hu2006-von}. Having redundant backup copies of primary objects, redundant network connections and redundant servers~\cite{hampel2006-p2pmmog} are ways to provide fault tolerance.

In games, one of the main concerns is cheating, which corresponds to \KEYW{security} in other applications. \citeN{yahyavi2013-p2pmmog} present three categories for cheating: Interrupting Information Dissemination, which includes premature disconnection, flooding of the network, replay attacks and the dropping of updates to peers; Illegal Game Actions, which includes tampering of end nodes of the network, falsifying identity and the use of computer enhanced data where human readings are expected; and Unauthorized Information Access, which includes the tampering of end nodes or network traffic analysis in order to gain access to privileged information. %
It is easier to build a game using a centralized architecture; if a \PtoP communication is used instead of a centralized approach, then dealing with cheating (security) and maintaining control over the game remains a challenge~\cite{yahyavi2013-p2pmmog,hu2006-von}. Having more control means that the architecture is easier to manage and maintain~\cite{yahyavi2013-p2pmmog}.

\section{Evaluating Properties Against the Domain of \IoT}
\label{section:evaluation}

\IoT can be discussed from two perspectives, `Internet' centric and `Thing' centric~\cite{gubbi2013-vision,atzori2010-iot}; leading to different architectures \ie centered on cloud computing (`remote'), or centered on the user (`local'), respectively~\cite{gubbi2013-vision}. 
A complication that {\MMOW}s do not currently have to deal with is that \IoT network architectures can be (partially) disconnected \eg MANets~\cite{lopez2011-resource}. If (partially) disconnect networks are present, interaction with the local environment is most likely still possible, contrary to remote access.

Because of the projected size of the \IoT, \KEYW{scalability} is expected to be a major issue and often mentioned in literature~\cite{atzori2010-iot,gubbi2013-vision,lopez2011-resource,miorandi2012-adhoc}. Literature is divided on whether a centralized cloud computing, decentralized \PtoP, or a hybrid architecture will achieve the needed scalability (through the addition of resources) for \IoT. Arguments for cloud computing are based on the advantages of abundant cloud storage and processing capabilities, while being tightly controlled for efficient energy usage and reliability~\cite{gubbi2013-vision}. %
Counter arguments state that decentralized \PtoP architectures show promise~\cite{gubbi2013-vision,atzori2010-iot} and that a centralized architecture cannot lead to a truly scalable solution~\cite{lopez2011-resource} %
\ie wanting to reap the ``seemingly endless amount of distributed computing resources and storage owned by various owners''~\cite{gubbi2013-vision}. 
If billions of devices will be soon connected to the Internet, producing a potentially quadratic number of interactions~\cite{yahyavi2013-p2pmmog}, then somehow \IoT needs to facilitate those interactions~\cite{perera2014-contextaware}.
Currently, many sensor network solutions are sense-only solutions, with many-to-one interactions \ie only a small fraction of solutions are sense-and-react applications (using actuators), with many-to-many interactions~\cite{mottola2011-wsn}. 
To achieve scalability through reducing consumption, there seems to be only one construct mentioned in \IoT literature. From the domain of {\WSAN}s, the construct of `clustering', for both physical devices and also objects present on those devices, has been suggested for scalability; either abstract regions are formed~\cite{mottola2011-wsn} or a few nodes are elected as `cluster heads' to act as decentralized authorities for each cluster~\cite{lopez2011-resource}. A problem with current clustering protocols being that mobility of nodes has hardly been considered~\cite{lopez2011-resource}.

To achieve the appropriate \KEYW{responsiveness}, \IoT is said to require two classes of traffic: the throughput and delay tolerant elastic traffic; and, the bandwidth and delay sensitive inelastic (real-time) traffic. These two types can be further discriminated into various levels of quality of service~\cite{gubbi2013-vision}. \citeN{perera2014-contextaware} refer to two different classes of traffic, namely `event driven' versus `time driven', which correlates to `event-triggered' and `periodic', respectively, in \WSAN literature~\cite{mottola2011-wsn}. \citeN{perera2014-contextaware} mention that ``real-time data processing is essential''.

``The synchronisation [sic] of data across the architecture'' (\ie \KEYW{consistency}) is envisioned to be a big challenge in \IoT~\cite{lopez2011-resource}. In WSAN literature, four approaches emerge on how to provide access to data: similar to a relational database, as remotely accessible variables or tuples, through mobile code, or by message passing~\cite{mottola2011-wsn}. 
A number of projects in \IoT use `participatory sensing', where sensors on people are used to obtain readings local to the user~\cite{gubbi2013-vision}. A difficultly in \IoT is that (partially) disconnected networks exist, so that ``there is a potential for various non-homogeneous copies of object data across the architecture'', making support for `one-copy' [primary copy] equivalence problematic \ie ``the value of all copies should be identical after a transaction''~\cite{lopez2011-resource}.

The aim of \IoT is to include many platforms and devices with various resource limitations \eg cloud server clusters, desktop computers, mobile devices, sensor networks and
everyday objects; access to resources is also through various network infrastructure, such as a wireless personal area network (\eg Bluetooth), wireless local area network (\eg Wi-Fi), wireless wide area network (\eg 2G and 3G networks), and satellite network (\eg GPS)~\cite{perera2014-contextaware}. 
Not all `things' in the \IoT have sufficient energy and processing power to do comprehensive data processing (\eg sensors in a \WSAN); in that case, data must be networked to more powerful devices and processed there~\cite{perera2014-contextaware}.
Because billions of \IoT devices will potentially be communicating with each other, \IoT research has noted the importance of optimizing computation in the various `layers' of the \IoT infrastructure~\cite{perera2014-contextaware} \ie \KEYW{resource utilization}. An example of this would be the ``notion of distributing computation in order to reduce the communication overhead, which is generally termed in-network processing or in-network computing''~\cite{gubbi2013-vision}. In participatory sensory, people are centric~\cite{gubbi2013-vision}, but this can be generalized in the sense that interactions for an object in the \IoT is highly dependent on the objects surroundings \eg presence of other objects or people~\cite{perera2014-contextaware}.

With the existence of partially or permanent disconnected networks in \IoT~\cite{lopez2011-resource}, robustness and fault tolerance (\ie \KEYW{availability}) will become fundamental research topics in \IoT~\cite{miorandi2012-adhoc}. 
Considering the `extremely large scale' of \IoT and the `high level of dynamism in the network', self-organization is suggested as a solution~\cite{miorandi2012-adhoc} \eg supporting merges and splits of the network~\cite{lopez2011-resource}. 
In WSAN literature, very little research has been done with respect to having mobile nodes in a network~\cite{mottola2011-wsn}. And, programming approaches for {\WSAN}s provide for only limited guarantees in the face of the various types of hardware faults~\cite{mottola2011-wsn}.

\citeN{perera2014-contextaware} state that the ``\IoT paradigm will intensify the challenges in \mbox{\KEYW{security}} and privacy''. Security is related to concepts such as authentication, privacy and integrity. For \IoT, cryptographic algorithms commonly used in authentication are typically problematic in devices with various resource limitations (\eg sensor networks), using large amounts of energy and bandwidth~\cite{atzori2010-iot}. Authentication usually involves identifying people, but in \IoT identities are associated with objects~\cite{miorandi2012-adhoc}. If authentication is to be possible in (partially) disconnected environments, the procedure will have to be possible locally~\cite{lopez2011-resource}. Privacy refers to the access of data related to an individual~\cite{miorandi2012-adhoc}. While surfing the Internet, individuals usually play an active role in their privacy, but in \IoT sensors are expected to collect information about individuals passively, without them actively using an \IoT service and without control over what information is being collected~\cite{atzori2010-iot,perera2014-contextaware}. Once the information is generated, it will most like be retained indefinitely, unless a mechanism is in place to allow for `digital forgetting'~\cite{atzori2010-iot}. The purpose of data integrity is to prevent adverse modification of data without detection~\cite{atzori2010-iot,lopez2011-resource}. A criticism with {\WSAN}s and thus \IoT, is that hardware components are easy to physically attack, because they are largely unattended and that if wireless communication is used, it can be eavesdropped~\cite{atzori2010-iot}. Security and privacy are very much open issues in \IoT~\cite{perera2014-contextaware,miorandi2012-adhoc}.

\section{{\MMOW}s and \IoT, Comparison of Key Properties}
\label{section:discussion}

To achieve \KEYW{scalability} through adding resources, cloud computing and \PtoP solutions are being considered in the domain of \IoT, but research in the domain of {\MMOW}s is more advanced, with running platforms utilizing partitioning schemes such as regionalization and replication in combination with interest management. 
To reduce consumption, clustering has been suggested in \IoT, for devices and objects present on those devices. Clustering for devices in \IoT can be equated to the partitioning techniques regions and shards in {\MMOW}s; and, clustering for objects in \IoT can be equated to interest management in {\MMOW}s~\cite{yahyavi2013-p2pmmog,lopez2011-resource}. For example,\citeN{lopez2011-resource} suggests using clusters determined through the use of context information, which is very similar to the existing \MMOW project called \mbox{\textit{Donnybrook}} using `interest sets'~\cite{yahyavi2013-p2pmmog}.
There seems to be an overarching opinion that a single architecture or middleware will enable \IoT, but there is a risk of fragmentation~\cite{miorandi2012-adhoc}. 
With many working on their own platform, one can speculate if \IoT will be a collection of clouds, also facing interoperability problems found in multicloud computing \eg the locking in of clients~\cite{singhal2013-multicloud}.
If \IoT does fragment and considering the advances in the domain of {\MMOW}s, this means solutions from the domain of {\MMOW}s (specifically the Metaverse; see \citeN{dionisio2013-metaverse}) could be exapted to the domain of \IoT.

In the \IoT literature consulted, there was no mention of the consistency and \KEYW{responsiveness} trade-off found in {\MMOW}s. Real-time data is one of the properties that ``distinguish virtual worlds [{\MMOW}s] from other distributed systems''~\cite{liu2012-melding}. 
Real-time traffic will not be required in all applications of \IoT, but those applications that require delay sensitive inelastic traffic would be in the same class as an \MMOW. 
Contrary to cloud-based {\MMOW}s, response times in \IoT can be kept lower, if interactions are kept in a local environment rather than with remote resources, due to reduced networking latency.

Also not in \IoT literature, was any mention of the performance (availability) and consistency restrictions, in a partitioned space \ie a distributed system.
A metric for \KEYW{consistency} in {\MMOW}s is `drift distance'~\cite{hu2006-von}; for a moving entity in a virtual world, the drift distance is, the absolute value of: the distance between the position of the entity locally and its position remotely. When dealing with the physical world, the drift distance could be calculated between the local position of the entity and its physical position. This means that given a physical entity with multiple virtual replicas (\ie multiple versions of reality)
the copy most similar to local physical entity is the one most consistent.

Until rather recently \MMOW architectures have had the luxury of only having to support the desktop class of devices, connected to a server cluster, with {\MMOW}s on mobile devices only recently becoming more common. In \IoT, efficient \KEYW{resource utilization} is still very much a challenge. Some sensors are able to produce a continuous stream of data and networking that data to more powerful devices or the cloud might outweigh the benefit, given the limited computing power of sensors and mobile devices. If faced with (partially) disconnected networks, the benefit of accessing local resources must be considered. %
If {\MMOW}s are to interface with \IoT, this means {\MMOW}s must also face (partially) disconnected networks. 
To allow for the use of local resources in a disconnected state, authority over local resources can be delegated to the local environment; a caveat being that, although {\MMOW}s have considered \PtoP or hybrid architectures, engines have typically been centralized clusters, rather than geographically distributed systems.

Similar to resource utilization, \IoT research has already taken into account (partially) disconnected networks, but if {\MMOW}s are to interface with \IoT, issues with \KEYW{availability} must be dealt with. If a shard is disconnected, delay tolerant networking must be used so that the shard can be merge when a connection is established~\cite{nevelsteenDRAFT-virtualworlddef,demeure2008-transhumance}.

Authentication is not mentioned as an issue for {\MMOW}s as long as a central authority is used. If decentralized \PtoP is used for either \IoT or an \MMOW, \KEYW{security} remains an issue. \citeN{miorandi2012-adhoc} state that \IoT needs to move away from centralized approaches, to a fully distributed and dynamic approach. 
Privacy in an \MMOW is similar to authentication, except that rather than trying to access private player data, attackers are trying to gain sensitive information pertaining to game state. Most game cheats are primarily against data integrity. Similar to \IoT, end nodes in an \MMOW cloud are also easily attacked and the networking tampered with. Premature disconnections and decentralized \PtoP for {\MMOW}s has a likeness to partially disconnected networks in \IoT. 
Missing from the \IoT discourse, is the issue of maintenance; if \IoT is to be enabled through a single platform, how will updates to the platform be managed?

\section{Case Studies of an \MMOW Interfacing With or Mediating \IoT}
\label{section:casestudies}

In previous sections, properties have been found in the domain of {\MMOW}s, that overlap with the domain of \IoT. To show that these two domains do indeed overlap, three different case studies are presented that combine {\MMOW}s with \IoT. The case studies include:
\begin{inparaenum}[(i)]
\item MediaSense, a middleware to handle communication between nodes of a distributed \IoT network;
\item Immersive Networking, a framework based on the MPEG-V standard, which outlines an architecture for interoperability between virtual worlds ({\MMOW}s) and the physical world (via \IoT); and 
\item the Metaverse as an aggregate of services from information systems, including \IoT.
\end{inparaenum}

\subsection{\MMOW Interfacing with a Distributed \IoT Middleware}

\begin{figure}[b]
	\centering
	\includegraphics[height=130pt,width=.96\textwidth]{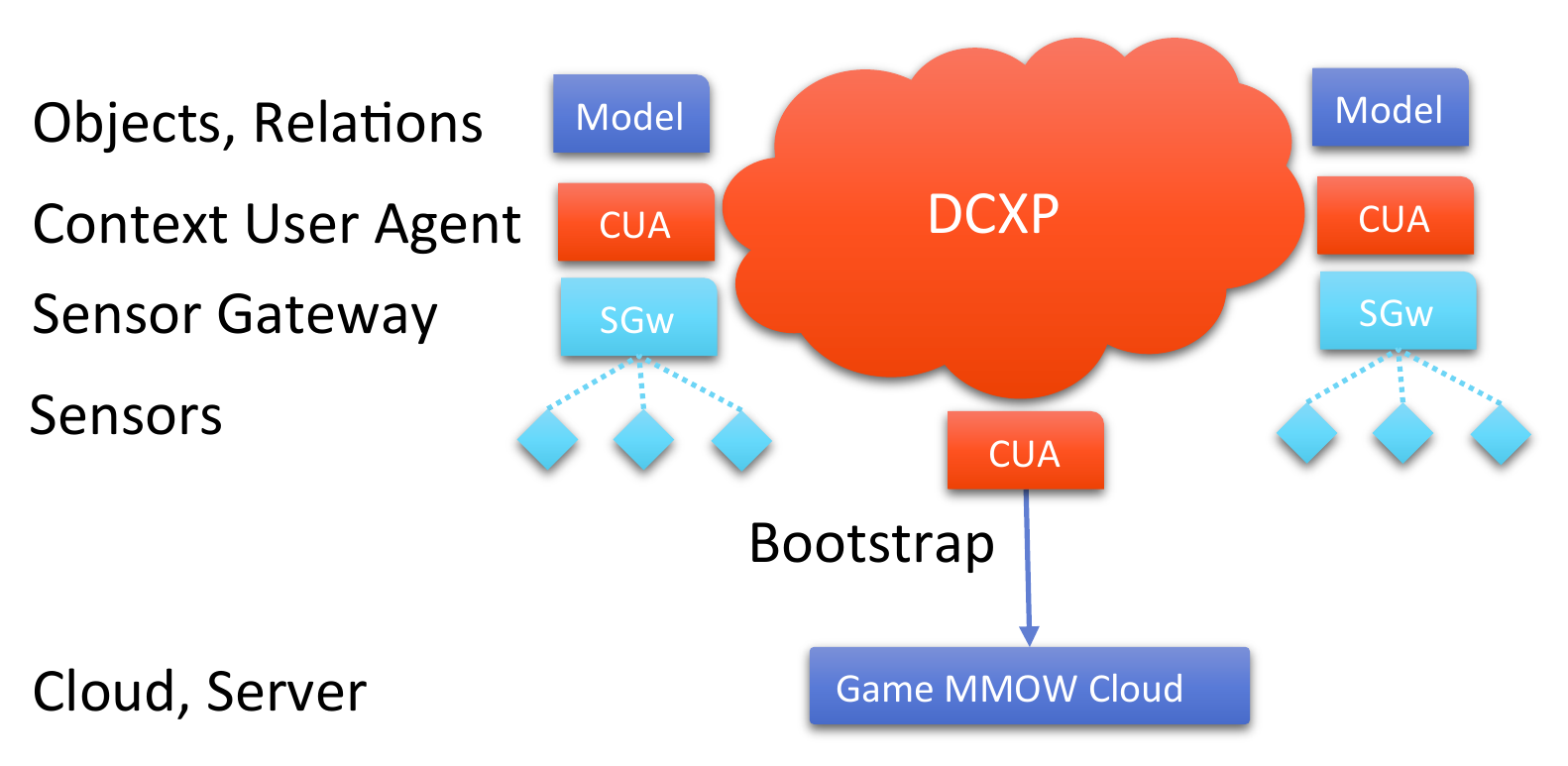}
	\caption{Connecting MMOW and a DCXP \PtoP context network}
	\label{figure:HotTown-Revisited}
\end{figure}

MediaSense~\cite{kanter2012-decentralized} is a distributed \IoT platform based on the Distributed Context eXchange Protocol (DCXP) for sharing context information between peers; locations of peers are denoted by Universal Context IDs and resolved via a Distributed Hash Table. 
Each end point in the middleware is referred to as a Context User Agent, which stores context information in the Distributed Hash Table as a \texttt{\{key,value\}} pair. Context information is modeled as relations between objects, which are maintained via DCXP methods (\texttt{GET/SET} or \texttt{PUBLISH/SUBSCRIBE}). 
The platform (see Figure~\ref{figure:HotTown-Revisited}) allows for a direct mapping of events between the virtual objects of a \MMOW by:
\begin{inparaenum}[(1)]
\item collocating a Context User Agent (running in a `bootstrap node'~\cite{kanter2009-dcxp} of the DCXP \PtoP context network) with a node from the \MMOW cloud;
\item creating an interface between collocated nodes so that virtual objects can be assigned to the Universal Context IDs of corresponding physical devices, on which a Context User Agent is running; and 
\item mapping game events (between virtual objects) to DCXP methods (between Context User Agents). %
\end{inparaenum}
New physical nodes can announce themselves and claim to represent a MMOW virtual object. 
Any node in the DCXP \PtoP context network can participate in the MMOW cloud; the only requirement being that at least one DCXP Context User Agent is collocated with the MMOW Cloud. Extrapolating the scenario means all objects are hosted in DCXP nodes, so that the game world (consisting objects and relations) is hosted by the DCXP \PtoP context network \ie effectively distributing the game engine. To achieve scalability for physical nodes, while maintaining relationships with other objects, \citeN{walters2014-thesis} presents a method for clustering objects according to multi-criteria ranking.

\subsection{\IoT Interfacing with Virtual Environments on a MMOW infrastructure}

MPEG-V~\cite{iso2009-mpegv} is a standard which outlines an architecture for interoperability between a virtual world ({\MMOW}s) and the physical world (via \IoT). In addition the standard aims to enable interoperability between virtual worlds by characterizing metadata for virtual world objects, so that objects can be migrated from one virtual world to another. MPEG-V has been drafted to support the \mbox{Metaverse} \ie forming a system of interoperable interconnected virtual worlds. %
Part~1 of the MPEG-V standard outlines a top level architecture, with target areas for standardization. 

\begin{figure}[!b]
	\centering
	\includegraphics[height=210pt,width=.96\textwidth]{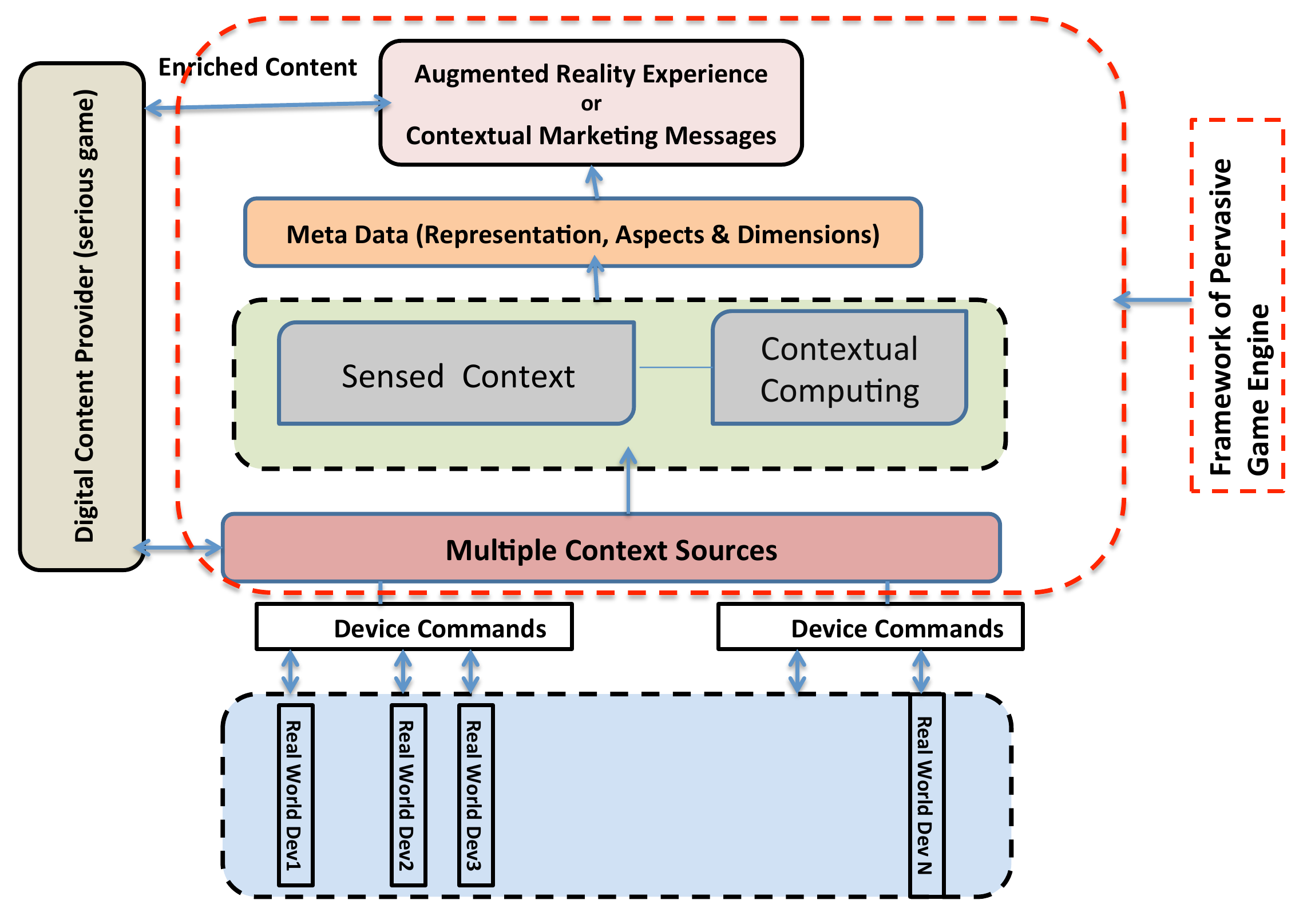}
	\caption[MPEG-V]{\textbf{Immersive Networking framework} - framework that aims to replace the adaptation engine (RV or VR) in MPEG-V standard}
	\label{figure:Figure-IoT-MMOW}
\end{figure}

Immersive Networking~\cite{kanter2016-immersivenet} aims to go beyond \IoT to enable the massive and scalable sharing of context information across a \MMOW infrastructure \ie information sharing in real-time for timely and intelligent decisions in different user scenarios. 
Immersive Networking is a framework technology that allows users to share context information in multiple practical scenarios to obtain immediate, objective feedback from the pervasive game engine towards a relational \IoT solution which is identified as interfacing with a virtual environment on a \MMOW infrastructure. 
The Immersive Networking framework is envisioned as a MPEG-V virtual world infrastructure with a distributed controller for interfacing with virtual environments, as depicted in Figure~\ref{figure:Figure-IoT-MMOW}.
The framework specifies the context and semantics required to provide interoperability in the distributed controlling of virtual agents as well as generic virtual objects. The framework aims to replace the adaptation engine (RV or VR) in MPEG-V,  making seamless experiences in virtual environments possible through the self-organization of connectivity between people, places and things.

\subsection{Virtual Worlds as a Behind the Scenes Resource}

\citeN{nevelsteen2015-pervasivemoo} concluded that a virtual world engine is in the same product line as a game engine for pervasive games. 
Considering pervasive games need a sensory system to monitor the physical world (through the use of non-standard input devices), IoT could potentially serve as such a sensory system. Since a \MMOW is a virtual world at a massive scale, this means that a \MMOW can be exploited as a `behind the scenes' resource for coordinating and managing devices and interaction in the physical space. 

Furthermore, according to Rehm et al.~\cite{rehm2015-mediator}, the concept of \IoT has been recently been replaced by the concept of Cyber-Physical Systems. Rehm et al. believe that ``virtual worlds can serve as platforms to facilitate the integration required by CPS [Cyber-Physical Systems]''. %
And, by extrapolation, they ``conceive of a unified platform, the Metaverse, built on VW [virtual world] technologies that allow for the integration of technological, physical, and human elements of CPS [Cyber-Physical Systems]''.

\section{Conclusions}
\label{section:conclusions}

The result of this article is the ``explication of the problem''~\cite{johannesson2014-designscience} of scaling \IoT, and incorporating a \MMOW with \IoT into a pervasive system.
Six properties, specific to a massive number of entities interacting, have been identified; first in the domain of {\MMOW}s, second in the domain of \IoT, and then compared in discussion. \IoT can clearly learn from advances, in availability with respect to \PtoP systems, and scalability, from the domain of {\MMOW}s. When virtual worlds start to incorporate \IoT, blending the virtual and the physical, {\MMOW}s can clearly learn from advances made, in resource utilization, availability, and responsiveness with respect to (partially) disconnected networks, from the domain of \IoT. For consistency and security, there seems to be advances in both domains that can cross over to the other domain. Rather than think that \IoT brings about a whole new set of issues, advances in other domains (\eg~the Metaverse) should be considered during \IoT development.

\section*{Author's contributions}
Research and drafting of property requirements relating Massive Multiplayer Online Worlds and the Internet of Things is done by Nevelsteen. Conceptual work on decentralizing a game engine for use with Internet of Things is done in collaboration with all three authors. Each of the case studies is provided by a single author; MediaSense by Kanter, Immersive Networking by Rahmani and Virtual Worlds as a ``behind the scenes'' resource by Nevelsteen.

\section*{Acknowledgements}
Research was made possible by a grant from the Swedish Governmental Agency for Innovation Systems to the Mobile Life Vinn Excellence Center.

\printbibliography[title=\textsc{References}]


\end{document}